\documentclass[prc,showpacs,twocolumn]{revtex4}

\usepackage{comment}
\usepackage{epsfig}
\usepackage{graphicx}
\usepackage{amsmath}
\usepackage{dcolumn}
\begin{document}

\title{Exit Doorway Model for Nuclear Breakup of Weakly Bound Projectiles}

\author{M. S. Hussein$^{1,2}$, R. Lichtenth\"aler$^2$ \\
\normalsize $^1${\it  Max-Planck-Instit\"ut f\"ur Physik komplexer Systeme, N\"othnitzer Str. 38, D-01187, Dresden, Germay} \\ 
\normalsize $^2$ {\it Instituto de Fisica, Universidade de S\~ao Paulo, C.P.
55318, 05315-970, S\~ao Paulo, SP, Brazil}}

\begin{abstract}
We derive closed expressions for the nuclear breakup cross sections in the adiabatic limit using the Austern-Blair theory. These expressions are appropriate for the breakup of weakly bound nuclei. The concept of an exit doorway  that mediates the coupling between the entrance channel and the breakup continuum is used.  We prove the validity of the scaling law that dictates that the nuclear breakup cross section scales linearly with the radius
of the target. We also compare our results for the nuclear breakup cross section of $^{11}$Be, $^8$B on several targets with recent CDCC calculation.

\end{abstract}

\maketitle

The breakup of nuclei is a common occurance when the bombarding energy is high 
enough and/or the binding energies are sufficiently low. In the case of weakly 
bound nuclei the threshold for breakup is small and more so for bound unstable 
nuclei. The mechanism of breakup is assumed to consist of  elongating the
 projectile, through the action of the interaction, which eventually leads to
 the production of two or more fragments. This interaction is composed of a 
short range, nuclear piece and a longer ranged electromagnetic one. A debate 
has been going on in the literature concerning the way the nuclear part of the
 breakup cross section depends on the mass of the target nucleus which supplies
 the interaction. In most references \cite{fuk,palit,aumann}, it is assumed 
that the dependence goes as the cubic root of the mass number. In reference 
\cite{nagara}, however, it is claimed that this dependence is more like linear!
In a recent paper \cite{tho}, through a careful Continuum Discretized Coupled 
Channels (CDCC) calculation, the former dependence ( A$^{1/3}$  ) has been 
established, which corroborates the contention that the nuclear breakup cross
 section should follow the prediction of the Serber model \cite{serber}.

It is interesting to compare the numerical CDCC calculation alluded to above 
with those of simpler analytical models. Specifically, the Ausern-Blair
 adiabatic theory for inelastic scattering comes to mind. If one assumes 
that the breakup proceeds through a so-called exit doorway 
\cite{hus,hus1,canto}, then the
 process can be treated as an inelastic excitation. The idea of exit doorway
 has been used in the case of the influence of breakup on fusion 
\cite{hus,hus1} 
and in 
the excitation of giant resonances \cite{canto} with success. In a very recent
 paper \cite{elis} a comparison of a preliminary CDCC calculation for the 
breakup cross 
section of the system $^6$He$+^{27}$Al at low bombarding energies with a simple formula derived by us]using the Austern-Blair model, showed that such an idea
 is quite reasonable and encouraged us to pursue the matter further. We do 
this in the present paper, where we fully develop the Austern-Blair model 
for the nuclear elastic breakup reaction cross section assumed to proceed through the excitation
 of an exit doorway \cite{hus,hus1,canto}.

The exit doorway concept has been used in the devolopment of reaction theories
 invloving the the excitation of a “doorway” in the final
state, in contrast to the conventional cases where such resonances are 
populated in the entrance channel \cite{auerbach}. In the breakup reactions 
of halo nuclei
one may envisage that the process proceeds through the breakup doorway
dipole, quadrupole etc.) into the continuum. As such, the detailed description
 of the exclusive reaction, where the final channels are especified, will 
necessarily contain the full information about the exit doorway ( its energy,
 width etc.). This is the case that was encountered in the theory of the
 excitation of multiple giant resonances \cite{canto} and of the influence 
of the
 pygmy resonance on the fusion of halo nuclei \cite{hus}. In the current 
paper we 
will be content with the inclusive quantity of the integrated breakup cross 
section and the only reference to the exit doorway is made implicitly as a
 final state that has to be populated for breakup to occur.

The full Hamiltonian which describes the colliding ions can be written as

\begin{equation}
H =H_0 + F     
\end{equation}

where $H_0 = h_0 + K + V = h_0+H^{(0)}$ is diagonal in open channel space, 
$h_0$  is the 
intrinsic part that describes the structure of the projectile and the target
 nuclei, 
$K$ is the kinetic energy operator  and $V$ is the optical potential which 
contains the complex nuclear plus the Coulomb parts. The operator $F$ describes
 the coupling among the open channels.

The intrinsic Hamiltonian $h_0$, which for simplicity is taken here to represent
the excitable projectile nucleus with the target considered structureless, is 
now written as:

\begin{eqnarray}
h_0=|\phi_0>E_0<\phi_0| +| d>E_d<d| + \sum_{i}|i>E_i<i| \nonumber \\ 
+  \sum_{i}[|d>\Delta_i<i| + |i>\Delta^*_i<d|] + \nonumber \\
\sum_{ij}[|i>\Omega_{ij}<j| + cc]                
\end{eqnarray}

The first three terms on the RHS above refer to the ground, exit doorway and 
discretized continuum states, respectively. The fourth term couples the doorway
 to the discretized continuum states, and the last term represents the 
continuum-continuum coupling. If we remove the doorway from the above 
we get the Continuum Disctretized Coupled Channels ( CDCC)
 intrinsic Hamiltonian

\begin{eqnarray}
h_0 = |\phi_0>E_0<\phi_0| + \sum_{i}|i>E_i<i| + \nonumber \\
 \sum_{i}[|\phi_0>\Delta_i<i| + |i>\Delta^*_i<\phi_0|] + \nonumber \\
\sum_{ij}[|i>\Omega_{ij}<j| + cc] 
\end{eqnarray}

The exit doorway modulated CDCC Hamiltonian, Eq. (2), is our subject of study 
here. A full development of this new CDCC will be left for a future
work. Here we concentrate our effort on understanding the consequence of 
reaching the breakup continuum from the entrance channel only through the exit
 doorway $|d>$. For this purpose we ignore the last term in Eq. (2) and
remind ourselves that, whereas $|\phi_0>$ and $|i>$ are eigenstates of $h_0$, 
$|d>$ is not.

The full doorway-modulated CDCC equations can be obtained as follows.
The full Schrodinger equation of the colliding system is,
\begin{equation}
[E-(H_0+F)]|\psi>=0                                                     
\end{equation}
which when projecting onto the different channels gives:
\begin{equation}
(E-E_0-H_0^{(0)})\psi_0^+=\sum_{i}F_{0i}\psi_i^+                             
\end{equation}
\begin{equation}
(E-E_i-H_i^{(0)})\psi_i^+=F_{i0}\psi_0^+                                       
\end{equation}

We now invoke the “ exit-doorway” hypothesis,

\begin{eqnarray}
 F_{0i}=F_{0d}\alpha^*_{di} & F_{i0}=F_{d0}\alpha^*_{id}
\end{eqnarray}

The overlaps $ \alpha_{di}$ and $\alpha^*_{id}$ 
and  can be easily obtained from Eq.(2)(without the last term)\cite{hus,canto},
\begin{equation}
|\alpha_{di}|^2=(\Gamma^{\downarrow}/2\pi)/[(E_i-E_d)^2+(\Gamma_d^{\downarrow}/2)^2] \label{alfa}
\end{equation}

where $\Gamma^{\downarrow}$ the exit-doorway “ spreading width “ describing 
its average coupling to the continuum states of the projectile, is related to
 the
 $\Delta_i$ factors
through,
\begin{equation}
\Gamma^{\downarrow}=2\pi\overline{|\Delta_i|^2}\rho
\end{equation}
where $\overline{|\Delta_i|^2}$ is an average value and $\rho$ is the average density of 
discretized continuum sates in the vicinity of d. Clearly the need to the 
continuum-continuum coupling terms would be very important if exclusive cross 
sections are to be calculated, since through them ( and through the doorway)
 the elastic channel´s coupling to the breakup channel continuum can be fully acounted for. Including the C-C coupling term, would result in a more complicated expression for $|\alpha_{di}|^2$ than that of Eq.\ref{alfa}.

Equations (5) and (6) can be recast into the following, after setting 
$E_0=0$ and $F_{ij}=0$,
\begin{equation}
(E-H_0^{(0)})\psi_0^{(+)}=\alpha_{di}F_{0d}\psi_i^{(+)}
\end{equation}
\begin{equation}
(E-E_i-H_i^{(0)})\psi_i^{(+)}=\alpha_{id}^*F_{d0}\psi_0^{(+)}
\end{equation}

The breakup cross section, within the exit doorway model then becomes,
\begin{eqnarray}
\sigma_{bu}=\sum_i\frac{k_i}{k_0}|\alpha_{di}|^2|<\psi_i^{(-)}|F_{d0}|\psi_0^{(+)}>|^2 & & \\
\approx \frac{|k_d|}{k_0}|<\psi_d^{(-)}|F_{d0}|\psi_0^{(+)}>|^2 & &
\end{eqnarray}

Where the sum over $i$ has 
been performed by appropriate contour integration over $E_i$.
Note that the Q-value in $\psi_d^{(-)}$ is complex owing to the non-zero width
of the exit dooway whose energy is $E_d-i \Gamma_d^{\downarrow}/2$. A simple way  to see how the complex Q-value arises is
 to eliminate  $\psi_i^{(+)}$ in Eq.(5) in favor 
of   $\psi_0^{(+)}$ by employing Eq.(6), which gives 
$\psi_i^{(+)} =\frac{1}{E-E_i-H_i^{(0)}+i\epsilon}F_{i0}\psi_0^{(+)}$.
With this Eq.(5) becomes $(E-E_0-H_0^{(0)}-\sum_i F_{0i}\frac{1}{E-E_i-H_i^{(0)}+i\epsilon}F_{i0})\psi_0^{(+)}=0$.
With the exit doorway hypothesis, the polarization potential contribution, 
$\sum_i F_{0i}\frac{1}{E-E_i-H_i^{(0)}+i\epsilon}F_{i0}$ becomes
\begin{eqnarray*}
\sum_i F_{0d}\frac{\Gamma^{\downarrow}/2\pi}{(E_i-E_d)^2+(\Gamma_d^{\downarrow}/2)^2}\frac{1}{E-E_i-H_i^{(0)}+i\epsilon}F_{d0} & & \\
\approx F_{0d}\frac{1}{E-(E_d-i\Gamma_d^{\downarrow}/2)-H_d^{(0)}+i\epsilon}F_{d0} & &
\end{eqnarray*}
This suggests defining the exit-dorway scattering wave function by setting
 $H_i^{(0)}=H_d^{(0)}$ such that Eqs.(10) and (11) become:
\begin{equation}
(E-H_0^{(0)})\psi_0^{(+)}=F_{0d}\psi_d^{(+)}
\end{equation}
\begin{equation}
(E-(E_d-i\Gamma_d^{\downarrow}/2)-H_d^{(0)})\psi_d^{(+)}=F_{d0}\psi_0^{(+)}
\end{equation}
The ``inelastic'' cross-section is thus given by Eq.(13) above with the aforementioned proviso that the Q-value 
of the excited state is complex. 
The width of this Q-value is a measure of the continuum contribution to the coupling.

At this point we comment on the inclusion of the continuum-continuum coupling, 
namely the last term in Eq.(3). In this situation the amplitudes 
$\alpha_{di}$ are obtained by matrix diagonalization and, among other things,
the resulting overlap probability $|\alpha_{di}|^2$, deviates from the Breit-Wigner form of Eq.(8). A possible form which may incorporate some of the c-c effects is  a Lorentzian: 
\begin{equation}
|\alpha_{di}|^2=\frac{2}{\pi}\frac{\Gamma^{\downarrow }_d E_i^2}
{(E_i^2-E_d^2)+\Gamma_d^{\downarrow 2} E_i^2} \label{alfa1} \nonumber
\end{equation}
The above form results in an equation for $\psi_d^{(+)}$ with a modified form
 factor  which depends on the position and width of the exit doorway
$\tilde{V}_{d0} \approx f(E_d,\Gamma^\downarrow_d)V_{d0}$ where 
$f(E_d,\Gamma^\downarrow_d)$ is generally complex. 
Accordingly the cross-section would be:
$\sigma=|f(E_d,\Gamma^\downarrow_d)|^2\sigma_{DWBA}$. In the limiting case of 
$\Gamma_d \ll E_d$, the factor  $f(E_d,\Gamma^\downarrow_d)$ is approximately given by
$(1-\frac{i\Gamma^\downarrow_d}{2E_d})^{1/2}$ resulting a cross-section given by:
$\sigma \approx (1+\frac{\Gamma^\downarrow_d}{2E_d})\sigma_{DWBA}$.
In the case of coupling to the breakup continuum considered here, the other limit, $\Gamma^\downarrow_d \gg E_d$ is more appropriate, as $E_d$ is roughly given by the Q-value of the breakup ($\leq 1$MeV) while $\Gamma^\downarrow_d$ measures the extent in continuum excitation the discretization is performed ($\approx 10$ MeV).
The function $f(E_d,\Gamma^\downarrow_d)$ can be calculated in such a situation, but we leave this for a future investigation. The important point we are making here, is that a DWBA calculation with complex excitation energy in the final state, and with a form factor of the type $f(E_d,\Gamma^\downarrow_d)V_{d0}$, should be an adequate candidate to treat the elastic breakup process.

In the following we take the exit doorway to be excited sates of different multipolarities and use the Austern-Blair sudden/adiabatic 
theory \cite{austern,frahn}. We employ the Distorted Wave Born Approximation for $\psi_0^{(+)}$ and $\psi_d^{(-)}$.


The elastic breakup cross section and its dependence on the target masss can 
be analysed within the Distorted Wave Born Approximation (DWBA).
If we treat the breakup as an inelastic multipole process, the 
amplitude $T_{LM}=<\psi_d^{(-)}|F_{d0}|\psi_0^{(+)}>$ would look like:

\begin{eqnarray}
 T_{LM}=\sum_{l_f}(2l_f+1)^{1/2}(i)^{l_i-l_f}<l_fL;00|l_i0> \nonumber \\
<l_fL;-MM|l_i0>R^L_{l_f,l_i}(k_f,k_i) \nonumber \\
e^{i\sigma_{l_f}(k_f)+\sigma_{l_i}(k_i)} 
Y_{l_f,-M}(\theta,0).
\end{eqnarray}  

The unpolarized cross section of the dipole transition is then obtained from the expression

\begin{equation}
d\sigma_{L}/d\Omega=\frac{\mu}{(2\pi\hbar^2)^2}\frac{k_f}{k_i}
\sum_{M=-L}^{M=L}|T_{LM}|^2 
\end{equation} 

The radial integrals $R^L_{l_f,l_i}(k_f,k_i)$ for pure nuclear excitation are  given by,
\begin{equation}
R^L_{l_f,l_i}(k_f,k_i)=(4\pi/k_fk_i)\int dr f_{l_f}(k_f,r)F_L(r)f_{l_i}(k_i,r)
\end{equation}
Where the form factors $F_L(r)$ are given by the following expressions for the monopole,$ L =0$, dipole, $L= 1$ and quadrupole excitations $L=2$ 
\cite{satchler},

\begin{eqnarray}
F_0(r)=-\delta_0^{(N)}[3 V(r)+r\frac{dV(r)}{dr}] & & \\
F_1(r)=-\delta_1^{(N)}(\frac{3}{2}) (\frac{\Delta R_P}{R_P})[\frac{dV(r)}{dr}+(\frac{R}{3})\frac{d^2V(r)}{dr^2} ] & & \\
F_2(r)=-\delta_2^{(N)}\frac{dV(r)}{dr} & & 
\end{eqnarray}                                                            
with $\Delta R_P = Rn - Rp$ being the difference between the rms radii of the neutron and proton distributions of the projectile, and $V(r)$ is the elastic scattering channel 
optical potential. The quantities $Rn$ and $Rp$ can be extracted from the 
analysis of Refs.\cite{alkha,kraszna}. In Ref.\cite{kraszna} a power expansion
 in $\Delta R_P$ was employed in the analysis of $\alpha$-inelastic scattering 
from neutron skin nuclei.

In the adiabatic limit, $k_i = k_f = k$, and for large orbital angular 
momenta, $l_f = l_i =l$, the radial integral can then be evaluated in closed 
form following the procedure of Austern and Blair \cite{austern,frahn}. 
For the dipole and quadrupole cases we have
\begin{eqnarray}
R^{(1)}_{l,l}(k)=-i\delta^{(N)}_1(\pi\hbar^2/\mu)(\frac{3}{2})(\frac{\Delta R_P}{R_P}) \times & & \nonumber \\
  (\frac{dS^{(N)}_l(k)}{dl}+(\frac{R}{3})\frac{d^2S^{(N)}_l(k)}{dl^2}) & & \\ 
R^{(2)}_{l,l}(k)=-i\delta^{(N)}_2(\pi\hbar^2/\mu)\frac{dS^{(N)}(k)}{dl} & &
\end{eqnarray} 

where $\delta^{(N)}_L$ is the nuclear deformation length given by 
 $\delta^{(N)}_L=\beta^{(N)}_L R_P$ with $\beta^{(N)}_L$ being the nuclear 
deformation parameter and $R_P$ is the radius of the excited projectile.

The above expression for the radial integrals can be associated with the nuclear elastic breakup radial integral. Thus we can obtain
analytical expression for the integrated nuclear breakup cross section
by simply integrating the cross section formula, eq.(2). In performing this calculation the angular momentum coupling coefficients are evaluated exactly and the sum over $l_i$ can be performed by putting the Coulomb phase shifts both as 
functions of $l_f \equiv l$. 
The amplitude of eq. (13) is given now by
\begin{equation}
T_{LM}=i\sqrt{2}\sum_{l=0}^\infty(2l+1)^{1/2}R^L_{l,l}(k)
e^{2i\sigma_{l}(k)} 
Y_{l,-M}(\theta,0).
\end{equation}       
with the condition that $T_{LM}=0$ if $L+M$ is odd.
The integrated pure nuclear breakup cross section containing dipole and 
quadrupole contributions then becomes the following

\begin{equation}
\sigma=[(\delta^{(N)}_1)^2(\frac{3}{2})^2(\frac{\Delta R_P}{R_P})^2+(\delta^{(N)}_2)^2] 
\sum_{l=0}^\infty(2l+1)|\frac{dS^{(N)}(k)}{dl}|^2
\end{equation}
 where terms proportional to the second derivative of $S^{(N)}_l (k)$ 
have been dropped.

A simple estimate of the above formula can be made by approximating the 
sum in $l$ by an integral in $\lambda=l+1/2$:
\begin{equation}
\sum_{l=0}^\infty(2l+1)|\frac{dS^{(N)}(k)}{dl}|^2 \rightarrow \int_0^\infty 
2\lambda|\frac{dS^{(N)}(k)}{d \lambda}|^2 d \lambda=I \label{I}.
\end{equation}

Assuming a real nuclear S-matrix  which depends on $\lambda$ 
through $[1+\exp(\lambda-\Lambda)/\Delta]^{-1}$ then the derivative of 
$S$ would peak 
around the grazing angular momentum $\Lambda$ with a width given by $\Delta$.  
The integral~(\ref{I}) is then obtained as: 
$I= \frac{\Lambda}{3\Delta}$ for $\Lambda/\Delta>>1$.
Using $\Lambda=kR$
$\Delta=ka$, with $R=r_0(A_p^{1/3}+A_t^{1/3})$ and $a$ being the diffuseness 
of the optical potential we find the simple formula for $\sigma$.

\begin{equation}
\sigma=c[(\delta^{(N)}_1)^2(\frac{3}{2})^2(\frac{\Delta R_P}{R_P})^2+(\delta^{(N)}_2)^2]
\frac{R}{3a} \label{AB}
\end{equation} 
where $c$ is a constant normalization factor which depends among other things on the exit doorway nature of the excited state exemplified by the factor
 $f(E_d,\Gamma\downarrow_d)$. 
It is clear that $\sigma$ depends linearly
on the radius of the target and, more importantly on the square of the nuclear 
dipole and quadrupole deformation lengths. Thus, the $A_T^{1/3}-$dependence
is established. 

In the calculation to follow we use the cluster model to calculate de deformation lengths for the different multipolarities \cite{bert1,bert2,bert3}
This model assumes that the projectile is composed of two clusters, a core
of mass and charge $a_c$ and $z_c$ and a ``valence'' particle with $a_b$ and 
$z_b$. The separation energy is denoted by $Q$, the Q-value of the breakup.
Calling the spectroscopic factor of finding the cluster configuration in the 
ground state of the projectile, $S$ one obtains the following expression for the distribution of $B(E \lambda)$ in the excitation energy $E_x$ 
\cite{bert1,bert3}.
\begin{eqnarray}
\frac{dB(E\lambda)}{dE_x}=SN_0^2\frac{2^{\lambda-1}}{\pi ^2}(\lambda!)^2(2\lambda+1)(\frac{\hbar^2}{\mu_{cb}})^\lambda \times \nonumber \\
\frac{Q^{1/2}(E_x-Q)^{\lambda+1/2}}
{E_x^{2\lambda+2}}\times \nonumber \\ 
(\frac{Z_bA_c^\lambda+(-1)^\lambda Z_c A_b^\lambda}{A_p^\lambda})^2e^2 \label{B}
\end{eqnarray}

where $N_0$ is normalization factor which takes into account the finite range,
$r_0$ of the $c+b$ potential. The latter is assumed to be such as to give a Yukawa type wave function at large distances,
$\psi_{bc}(r)=N_0 \sqrt{K/(2\pi)}\frac{e^{-Kr}}{r}$ with 
$K=\sqrt{2\mu_{bc}Q/\hbar^2}$ and $N_0=\frac{e^{Kr_0}}{\sqrt{1+Kr_0}}$.
It is easy to obtain $B(E\lambda)$ by simply integrating of Eq.(\ref{B}) and employing the expression:

\begin{equation}
\int_0^\infty\frac{y^{\lambda+\frac{1}{2}}}{(y+1)^{2\lambda+2}}dy=
\frac{(-)^{2\lambda+3}\pi}{(2\lambda+1)!sin[(\lambda+\frac{3}{2})\pi]}
\prod_{k=1}^{2\lambda+1}(\lambda+\frac{3}{2}-k)
\end{equation}

We get for the cluster-model deformation lenghts $\delta_1^2$ and 
$\delta_2^2$ the following:
\begin{eqnarray}
(\delta^{(N)}_1)^2=(\frac{2\pi}{3}\frac{A_P}{Z_PN_P})^2\frac{B(E1)}{e^2}=& &\nonumber\\
=N_0^2S(\frac{2 \pi A_P}{3Z_PN_P})^2\frac{3}{16\pi}\frac{\hbar^2}{\mu_{bc}}
(\frac{A_cZ_b-A_bZ_c}{A_P})^2\frac{1}{Q} \label{d1}
\end{eqnarray}

\begin{eqnarray}
(\delta^{(N)}_2)^2=(\frac{4\pi}{3Z_PR_P})^2\frac{B(E2)}{e^2}=\nonumber\\
=N_0^2S(\frac{4 \pi A_P}{3Z_PR_P})^2\frac{5}{32\pi}\left(\frac{\hbar^2}{\mu_{bc}}\right)^2
(\frac{A_c^2Z_b+A_b^2Z_c}{A_P^2})^2\frac{1}{Q^2} \label{d2}
\end{eqnarray}

where $p(=b+c)$ refers to the projectile.

For our three nuclei discussed here, we have $^{11}$Be$=^{10}$Be$+n$,
 $^{8}$B$=^{7}$Be$+p$ and  $^{7}$Be$=^{4}$He$+^{3}$He, which define their cluster character, with the corresponding breakup Q-values, $0.504$ MeV, $0.137$ MeV and $1.587$ MeV. The factor $N_0^2S$ could be related to the Asymptotic Normalization Coefficient (ANC) of the bound state wave function and is taken as a parameter to be adjusted so as to account for the experimentally known $B(E\lambda)$.

\begin{figure}
\includegraphics[width=0.5\textwidth]{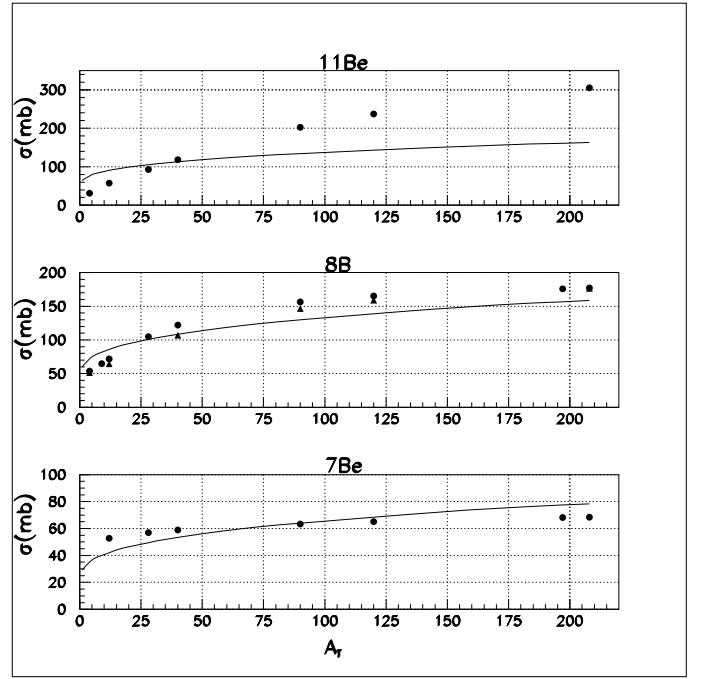}
\caption{\label{Fig}CDCC calculations for the nuclear breakup (dots) compared to the results of Eq.~(\ref{AB}). See text for details.}
\end{figure}

Simple estimate of $\frac{\Delta R_P}{R_P}$ can be obtained from 
\cite{satchler,satchler1} who gave 
$\frac{\Delta R_P}{R_P} \approx \frac{|N_p^{1/3}-Z_p^{1/3}|}{A_p^{1/3}}$.
In Table 1 we present the results of the deformation lengths obtained using the cluster model of formulas \ref{d1} and \ref{d2}.
For $^{11}$Be, we used $B(E1)=1.05 \pm 0.06$ e$^2$fm$^2$ \cite{fuk} and we get
$(\delta^{(N)}_1)^2=0.71\pm0.04$ fm$^2$. 
Further,$(\delta^{(N)}_2)^2=1.27\pm0.25$ fm$^2$ from the same reference. 


In figure 1 we compare our results, Eq.(\ref{AB}) using $a=0.65$ fm with the CDCC calculation of Ref.\cite{tho} at $E_{lab}=200$ MeV.A.
Clearly we underestimate the CDCC calculation. The reason resides in the neglect, in our model, of the higher-order channel 
coupling terms alluded to above. We also show in figure 1 the comparison with the CDCC calculation for $^8$B and $^7$Be 
using the formula(\ref{AB}). Clearly the scaling law is better 
obeyed in the ``normal'' nucleus 
$^7$Be as has already been discussed in \cite{tho}. The value of the normalization $c$ is close to unity for the ``normal'' nuclei $^7$Be
For the $^8$B the normalization is very close
the average value of the Asymptotic Normalization Coefficient
$ANC=0.45$ measured in ref.\cite{trache}.
For the $^{11}$Be a higher normalization is obtained probably due to higher 
order effects which are not accounted by our DWBA description. 
\begin{table}
\caption{Deformation lengths for the $^7$Be, $^8$B and $^{11}$Be projectiles. 
The deformation lengths for $^7$Be and $^{8}$B have been calculated
using formulas (\ref{d1}) and (\ref{d2}) using $N_0S=1.$
For the $^{11}$Be the $\delta_1 $ and $\delta_2$ are the values from 
Ref.\cite{fuk}} 
\begin{tabular}{|c|c|c|c|c|}
\hline
Projetile & $(\delta_1^{(N)})^2$(fm$^2$) &$(\delta_2^{(N)})^2$  &  $(\frac{\Delta R}{R})^2$ & $c$ \\
\hline
$^7$Be  &  0.41 &  1.54     &   0.00576   &   1.37 \\
\hline
$^{8}$B  &  1.62 &  3.24      &   0.0179  &   0.307 \\
\hline
$^{11}$Be  &  0.84(2) &  1.27(25)  &  0.0214  & 2.52\\
\hline
\end{tabular}
\end{table}

In conclusion, we derived an expression for the nuclear breakup cross-section using the Austern-Blair theory. 
The obtained cross-section exhibits the scaling 
law and should serve to supply a simple mean for an estimate of the nuclear breakup contribution.
The expression found should be contrasted with the purely geometric Serber-like expression $\sigma=2 \pi Ra$\cite{serber}.

\section*{Acknowledgements}

This work is supported in part by Funda\c{c}\~ao de Amparo \`a Pesquisa do 
Estado de S\~ao Paulo (FAPESP) and the Conselho Nacional de Desenvolvimento 
Cient\'{\i}fico e Tecnol\'ogico (CNPq). M.S.H is the Martin Gutzwiller Fellow,
2007/2008.

\end{document}